\input{aipcheck}

\documentclass[
    ,final            
  ]
  {aipproc}

\layoutstyle{6x9}

\newcommand{\ba}{\begin{eqnarray}}
\newcommand{\ea}{\end{eqnarray}}

\newcommand{\beq}{\begin{equation}}
\newcommand{\eeq}{\end{equation}}
\newcommand{\beqs}{\begin{eqnarray}}
\newcommand{\eeqs}{\end{eqnarray}}

\newfont{\prg}{cmsy10}

\begin{document}


\title{Equivalence Principle 
  and Partition \\ of Angular Momenta in the Nucleon }

\classification{11.80.Cr, 12.40.Nn, 13.85.Dz}
\keywords      {spin, gravity}

\author{O.V. Teryaev   \thanks{teryaev@theor.jinr.ru} } {
  address={\it BLTP, \\
Joint Institute for Nuclear Research,
141980 Dubna, Moscow region, Russia}
}



\begin{abstract}
 The manifestation of equivalence principle (EP) in spin-gravity
 interactions,resulting in the nullification of the corresponding analog of Anomalous Magnetic moment is
 explored. Its tests in the experiments with atoms and cold neutrons are
 discussed. The validity of EP separately for quarks and gluons in
 the nucleon resulting in exact equipartition of momentum
 and total angular momentum is conjectured. The important role of
 relocalization (Belinfante) invariance in these and other aspects
 of nucleon spin structure is stressed.
\end{abstract}

\maketitle



Equivalence principle is known to be one of the basic postulates
of the modern physics, constituting the cornerstone of General
Relativity. Its simplest and well-known "Newtonian" counterpart
corresponds to the equality of inertial and gravitational mass and
is tested with good accuracy. At the same time, there is another,
"post-Newtonian", manifestation of equivalence
principle 
which corresponds to the interaction of {\it spin with gravity}
\cite{KO}.
It means the absence of 
gravitational analogs of electric dipole and anomalous magnetic
moments. It may be derived as a low energy theorem due to the
conservation of momentum and orbital angular momentum.
As soon as these conservation laws control (due to Ji sum rules
for Generalized Parton Distributions) the partition of momentum
and angular momentum between quarks and gluons, the equivalence
principle is manifested in this context\cite{T1}. The connection
is provided by gravitational formfactors, being the  matrix
elements of Belinfante energy-momentum tensors, and in, turn, to
the total angular momenta of partons,
\begin{eqnarray}
      \langle p'| T_{q,g}^{\mu\nu} |p\rangle
       &=& \bar u(p') \Big[A_{q,g}(\Delta^2)
       \gamma^{(\mu} p^{\nu)} +
   B_{q,g}(\Delta^2) P^{(\mu} i\sigma^{\nu)\alpha}\Delta_\alpha/2M ] u(p),
\label{def}
\end{eqnarray}
where $P^\mu=(p^\mu+{p^\mu}')/2$, $\Delta^\mu = {p^\mu}'-p^\mu$,
and $u(p)$ is the nucleon spinor. We dropped here the irrelevant
terms of higher order in $\Delta$, as well as containing $g^{\mu
\nu}$. The parton momenta and total
angular momenta are:
\begin{eqnarray}
      P_{q, g} =  A_{q,g}(0), \nonumber \\
      J_{q, g} = {1\over 2} \left[A_{q,g}(0) + B_{q,g}(0)\right] \ .
\end{eqnarray}
Taking into account the conservation of momentum and angular
momentum one get
\begin{eqnarray}
\sum_{i=q,G} \int_0^1 dx x H_i (x,\xi,Q^2)=A_q(0)+A_g(0)= 1 \label{m} \\
\sum_{i=q,G} \int_0^1 dx x (H+E)_i
(x,\xi,Q^2)=A_q(0)+B_q(0)+A_g(0)+B_g(0)=1 \label{am},
\end{eqnarray}
which is just EP.
Note that gravitational analog of dipole moment is absent, as it
violates EP as well as CP invariance, while the obvious
gravitational analog of anapole moment of the form
$$\bar u(p')
\gamma^{(\alpha}  \gamma_5 p^{\nu)}( \Delta^2 g^{\mu\alpha}- \Delta_\alpha
       \Delta_\mu)
   u(p)
   $$
is allowed. The spin-dipole coupling appears at the level of
Hamiltonian entering the unitary transformed Dirac equation
\cite{O}, although it disappears \cite{PRD} when the corresponding
transformation of dynamical variables is properly considered.

Note that these formfactors describe also the interaction of
nucleons with TeV scale extra-dimensional gravity and they should
be taken into account when respective gravitational effects in
diffractive scattering \cite{selugin} are considered.

The dedicated test of the nullification of total Anomalous
Gravitomagnetic moment (AGM) $B$ was not yet performed. However,
there is a recent \footnote{I am indebted to J. Ellis for pointing
out this reference} claim \cite{com} that spin-rotation coupling
should be already taken into account when analyzing the data
obtained in the precise EDM experiment \cite{baker}. Moreover,
earlier atomic experiment \cite{venema} may be interpreted \cite{ST06} as
a test of EP with a few percent accuracy. Originally this
experiment was aimed on the search of gravitational dipole term,
but as it violates also CP invariance, the CP conserving EP
violating effects should be considered as a dominant ones,

Ultracold neutrons can also be used in interferometer experiments
with the rotating spin-flippers \cite{Mashexp} and implemented at
the existing and developed interferometers at ILL and Tokai
\cite{Private}. It seems reasonable to have two (rather than one
as suggested in \cite{Mashexp}) rotating spin-flippers. Their
rotation in the same and opposite directions may provide a number
of true and false signals. 
Namely, signals should be absent if they are rotated in the same
directions, and it should be twice larger in the case of rotation
in opposite direction in comparison yo the case when only one flipper rotates.
In total, there are 8 signals which may significantly increase the
statistics. Note also that EP tests may be performed 
in the experiments with polarized electrons and positrons in storage rings \cite{ST06}

There are also some evidences supporting the conjecture \cite{T2}
that EP is valid separately for quarks and gluons in the nucleon,
which is violated in perturbative QCD but may be restored in full
non perturbative (NP)QCD due to the phenomena of confinement and spontaneous chiral
symmetry breaking. This (extended) EP means exact equipartition
(EEP) of momenta and angular momenta in a nucleon.

The most precise numerical support is coming from the lattice
simulations \cite{phil} obtained after the conjecture \cite{T2}.
Another one is coming from the numerical relations between
anomalous moments and spin-averaged distributions of valence
quarks \cite{GP}. It is based on the adopted parametrizations of
GPD (see also \cite{Wak}) and the physical reason is nothing else
than EEP which holds in this parametrization.

As soon as EEP is related to properties of NP QCD one may expect
that it is valid also for other hadrons, in particular for vector
mesons. This is supported by QCD sum rules calculations \cite{SR}
of the anomalous magnetic moment of $\rho$ mesons. The results are
close to value $g=2$ which may be explained as a result of the
smallness of the analog of anomalous GPD $E$, related in turn, to
EEP. Note that the gluonic momentum calculated in similar
\cite{SR2} approach is sizable and one cannot attribute EEP merely
to absence of gluonic contributions, like in model calculations \cite{Wak}.
Of course, the direct QCD SR calculations of AGM in line with \cite{SR}
would be very interesting.

The generality of EEP should imply also its validity for the case
of hadrons substituted by currents, which would change the matrix
elements to 3-point correlators. This may be especially suitable
for lattice calculations, when this correlator may be considered
as an order parameter for confinement (chiral) phase transition,
when EEP may be violated in deconfined phase. If one consider EEP
in the case of tensor currents, the respective order parameter
looks quite symmetric. 

Let us also note that EEP may be supported by the conjecture \cite{BHS} 
relating Sivers functions and anomalous magnetic moments.
Let us suppose that this relation may be quantified as a proportionality between GPD $E$ and  
Sivers function. If so,  EP directly corresponds to Burkardt \cite{burk} sum rule. 
Furthermore, EEP is a natural counterpart of the recent conjecture \cite{BG}
about the smallness of gluon Sivers function, supported by COMPASS data \cite{Franco}.
 
EEP is related to the important property of Relocalization (Belinfante) Invariance (RI)
providing the possibility to perform a transformation of the densities of conserved 
charges and 
represent the total angular momentum in an "orbital' form
with Belinfante symmetrized energy momentum tensor (EMT). Let us stress once more that it is this tensor
which describes the coupling to gravity and enters the gravitational formfactors. 

The matching of RI with quantum theory happens not to be trivial.  
The analysis of leading order QCD evolution \cite{ot98} based on consistent exploration of conservation laws 
shows that it is RI
that leads to EEP due to relation between spin-dependent and spin-independent 
kernels
\begin{eqnarray}
\int_0^1 dx x \Delta P_{Gq}(x)=
{1\over 2} \int_0^1 dx x P_{Gq}(x).
\label{c}\end{eqnarray}
While RI for classical fields requires their decrease at infinity 
it puts the constraints for the behaviour of the matrix elements
of respective operators at low momentum transfers which are of special 
interest in 
In the non-perturbative case, when RI leads \cite{ot93} to the relation 
\begin{equation}
q^2{\partial \over{\partial q^{\alpha}}}
\langle P | J_{5S}^{\alpha}| P+q\rangle
=(q^{\beta}{\partial \over{\partial q^{\beta}}}-1)
q_{\gamma}\langle P | J_{5S}^{\gamma} |P+q \rangle,
\end{equation}
valid up to the terms linear in $q$ and excluding the possibility of massless pole in the matrix element of singlet axial current.
Thus RI provides a complementary view to such important property of NP QCD as $U_A(1)$problem. 

Note also that RI provides a guideline for dealing with non-local operators.
Indeed, matrix element of the contribution od antisymmetric part of quark EMT to 
angular momentum should 
be equal, due to IR, to that of quark spin. This requires that derivative 
resulting from $x$ factor in angular momentum should act just to matrix element,
picking the term linear in $\Delta$, while the singular coefficient is 
the same delta-function as for local operators. 
This provides the regular way of deriving the sum rules with no
need to use the wave packets like in the detailed analysis of \cite{BLT}.

The resulting sum rules for the longitudinal and transverse \cite{PST} polarization are completely similar. 
Say, the quark spin is described by the different projections of {\it chiral-even} axial current. 
The appearance of chiral-odd structures \cite{BLT} is due to the use of the 
partonic description in terms of wave functions, for which chiral even and odd structures are related.

To conclude, various experimental checks of equivalence principle in spin-gravity interactions
may be extracted from current and future experiments \cite{ST06}. 
The generalization of EP in NPQCD is supported by a number of observations
and may be related to Belinfante invariance. The possibility of deep roots 
of EEP in some NPQCD relations to gravity, say, to AdS/CFT correspondence, is 
very interesting.

I am indebted to Organizers for warm hospitality in Kyoto and financial support. 
This work was supported in part by Grants RFBR
06-02-16215 and RF MSE RNP 2.2.2.2.6546.




\bibliographystyle{aipproc}   


\bibliography{sample}

\IfFileExists{\jobname.bbl}{}
 {\typeout{}
  \typeout{******************************************}
  \typeout{** Please run "bibtex \jobname" to optain}
  \typeout{** the bibliography and then re-run LaTeX}
  \typeout{** twice to fix the references!}
  \typeout{******************************************}
  \typeout{}
 }

\end{document}